\documentclass[onecollarge,natbib]{svjour2}
\bibstyle{spbasic}

\bibpunct{[}{]}{;}{n}{}{,} 
\smartqed  
\usepackage{graphicx}

\newcommand{\ben}{\begin{displaymath}}
\newcommand{\een}{\end{displaymath}}
\newcommand{\be}{\begin{equation}}
\newcommand{\ee}{\end{equation}}
\newcommand{\bea}{\begin{eqnarray}}
\newcommand{\eea}{\end{eqnarray}}

\newcommand{\eq}[1]{Eq.~(\ref{#1})}

\newcommand{\bfp}{{\bf p}}

\journalname{Few-Body Systems}
\begin{document}

\title{Hidden Color and the $b_1$ structure function of the deuteron. 
}


\author{Gerald A. Miller   
}


\institute{Gerald A. Miller at
              Department of Physics, Univ. of Washington, Seattle, WA, 98195-1560 \\
              \email{miller@uw.edu}           
       }

\date{Received: date / Accepted: date}

\maketitle

\begin{abstract}
The $b_1$ structure function is an observable feature of a spin-1 system sensitive to non-nucleonic components of the target nuclear wave function.   A simple model for hidden-color, six-quark configurations   is proposed and found to give substantial contributions for values of $ x>0.2$. Good agreement with Hermes data is obtained. Predictions are made for an upcoming JLab experiment. 
\end{abstract}
  \def\quup{q^1_{\uparrow}}
\def\qudp{q^{1}_{\downarrow}}
\def\quum{q^{-1}_{\uparrow}}
\def\qudm{q^{-1}_{\downarrow}}
\def\quuz{q^0_{\uparrow}}
\def\qudz{q^0_{\downarrow}}
\section{Introduction}
\label{intro}
This document is 
based on the paper~\cite{Miller:2013hla}. That paper considered  several different effects that contribute to the $b_1$ structure function of the deuteron. Here  only  the effects of hidden-color components of the deuteron are discussed.

 Deep inelastic scattering from a spin-one target has features, residing in the leading-twist  $b_1$ structure function,  that are not  present for a spin-1/2 target~\cite{jaffe,frankfurt}.
  In the Quark-Parton model  
  \bea b_1= \sum_q e_q^2 \left[\quuz-{1\over2}(\quup+\quum)\right]\,\equiv\sum_ie_q^2\delta q_i\label{pm1}
  \eea
where $q^{m}_{\uparrow}$($q^{m}_{\downarrow}$) is the number density of 
quarks with spin up(down) along the $z$ axis in a target hadron with
helicity $m$.  The function $b_1$ is called the tensor structure function of the deuteron because it has been observed  by the Hermes collaboration using a tensor polarized deuteron target~\cite{Airapetian:2005cb} for  values of Bjorken $0.01<x<0.45$.  The function  $b_1$ takes on its largest value of  about 10$^{-2}$ at the lowest measured value of  $x$ (0.012),   decreases  with increasing $x$  through zero and takes on a minimum value of roughly $-4\times 10^{-3}$. 

The function $b_1$ nearly vanishes if the spin-one target is made of constituents in a relative $s$-state, and is very small for a target of spin 1/2 particles moving non-relativistically in higher angular momentum states~\cite{jaffe,Khan:1991qk,Umnikov:1996qv}.  Thus one expects~\cite{jaffe}  that a nuclear $b_1$  may be dominated by non-nucleonic components of  the target nuclear wave function. Consequently, a Jefferson Laboratory experiment~\cite{Jlab} is planned to measure $b_1$ for values of $x$ in the range 
$0.16<x<0.49$  and $1<Q^2<5 $ GeV$^2$ with the aim of reducing the error bars.  Our focus here is on the kinematic  region of the largest higher values of $x$ that are available to the JLab experiment. The Hermes experimental result~\cite{Airapetian:2005cb} presents an interesting puzzle because it observed a significant negative value of $b_1$ for $x=0.45$. At such   a value of $x$, any sea quark effect such as arising from double-scattering or virtual pions is completely negligible. Furthermore, the nucleonic contributions are computed to be very small~\cite{Khan:1991qk,Umnikov:1996qv}, so one must consider other possibilities. We therefore take up the possibility that the deuteron has a six-quark component that is orthogonal to two nucleons.
Such configurations are known to be dominated by the effects of so-called hidden-color states in which two color-octet baryons combine to form a color singlet~\cite{Harvey:1988nk}. Such configurations can be generated, for example, if two nucleons exchange a single gluon leading to a quantum fluctuation involving an color octet and  color anti-octet baryon.

 In particular,   a component of the deuteron in which all 6 quarks are in the same spatial wave function ($|6q\rangle$) can be expressed in terms on nucleon-nucleon $NN$, Delta-Delta $\Delta\Delta$ and hidden color components $CC$ as~\cite{Harvey:1988nk}:
\bea |6q\rangle=\sqrt{1/9}|N^2\rangle+\sqrt{4/45}|\Delta^2\rangle+\sqrt{4/5}|CC\rangle.\label{cc}\eea
This particular state has an 80\% probability of hidden color and only an 11\% probability to be a nucleon-nucleon configuration.  
 The 80 \% cited here is a purely algebraic  number that applies only for completely overlapping nucleons. 
The real question is the probability that the deuteron consists of 6 quarks are in the same spatial wave function, which is denoted  here as $P_{6q}$.  
 A recent review of hidden color phenomena is presented in~\cite{Bakker:2014cua}. In the following, the term  $|6q\rangle$ is  referred to interchangeably as either a six-quark or hidden color state.

The discovery of the EMC effect~\cite{Aubert:1983xm} caused researchers to consider the effects of such six-quark states~\cite{Carlson:1983fs}  in a variety of nuclear phenomena~\cite{Miller:1985un,Koch:1985y,Miller:1987ku}. Furthermore, the possible discovery of  such a state as a di-baryon resonance has drawn recent  interest~\cite{Bashkanov:2013cla}. Therefore we propose a model of a hidden-color six-quark components of the $s$ and $d$-states of the deuteron.  We also note that including a six-quark hidden color component of the deuteron does not lead to a conflict with the measured asymptotic $d$ to $s$ ratio of the deuteron~\cite{Guichon:1983hi}.
The EMC effect remains the only nuclear effect that has not been explained using conventional (non-quark) dynamics~\cite{Arneodo:1992wf,Geesaman:1995yd,Hen:2013oha}. 

\section{Hidden-color six-quark states} 
  
 We  investigate the  possible relevance of hidden-color six-quark states.
 There have been many models and attempts to provide definitive evidence  for the existence of such states. However, no  model has been unambiguously and uniquely verified by experiment. This is because conventional  nuclear theory makes no reference to  such states and yet successfully reproduces all known nuclear phenomena except for the EMC effect. 
 In this paper, we take the view that the HERMES observation at $x=0.452$ (their largest value of $x$) may provide another example requiring unconventional nuclear wave functions, and therefore may  offer an opportunity to finally learn something definite.  The key point is that nucleonic (and mesonic)  effects, as presently computed, offer much smaller (in magnitude) values of $b_1$ than found  by HERMES. A value of $x=0.452$ involves valence quarks. Since nucleons do not provide a mechanism, one is naturally encouraged to look at six quark hidden color components, which should have support at such values of $x$.
Since no definitive model exists,  it is sufficient to use the simplest of many possible models for the present exploratory calculation.

We proceed by assuming the existence of  a deuteron component  consisting of
  six non-relativistic quarks in an $S$-state. As stated in the Introduction, such a state has only a probability of 1/9 to be a nucleon-nucleon component, and is to a reasonable approximation a hidden color state, so we use the terminology six-quark, hidden color state.  Then we obtain  the corresponding  $d$ state component by  promoting any one of the quarks to a $d_{3/2}$-state. We define these states by combining
  5 $s$-state quarks into a spin 1/2 component, which couples with  either the $s_{1/2}$ or $d_{3/2}$ single-quark state to make a total angular momentum of 1.
 We therefore write the  wave functions of these states for a deuteron  of $J_z=H$  as
  \bea
  \psi_{j,l,H}(\bfp)=\sqrt{N_l}f_l(p)\sum_{m_s,m_j}{\cal Y}_{jlm_j}\langle jm_j,{1\over2}m_s\vert1H\rangle,\label{wf}
 \eea 
 where $l,j=s_{1/2}$ or $d_{3/2}$,  
  $N_l$ is a normalization constant chosen so that $\int\,d^3p\bar{\psi}_{j,l,H}(\bfp)\gamma^+\psi_{j,l,H}(\bfp)=1$ and ${\cal Y}_{jlm_j}$ is a spinor spherical harmonic. The  matrix element  for transition between the $l=0$ and $l=2$ states is given by the light-cone distribution:
\bea &F_H(x_{6q})={1\over2}\int d^3p\bar{\psi}_{1/2,0,H}(\bfp)\gamma^+\psi_{3/2,2,H}(\bfp)\delta\left({p\cos\theta+E(p)\over M_{6q}}-x_{6q}\right),
\eea
where $E(p)=\sqrt{p^2+m^2}$ with $m$ as the quark mass, and $M_{6q}$ is the mass of the six-quark bag, $x_{6q}$ is the momentum fraction of the six-quark bag carried by a single quark and $x_{6q}M_{6q}=x M$~\cite{Carlson:1983fs}. Note that
  $p\cos\theta$ is the third ($z$)  component of the momentum, so that the plus component of the quark momentum is $E(p)+p\cos{\theta}$. 
   We take $M_{6q}=2M$ (its lowest possible value) to make a conservative estimate.
  
 The term of interest $b_1(x)$ is given by
 \bea b_1^{6q}(x)={1\over2}(2)\left(F_0(x)-F_1(x)\right)P_{6q},\eea
 where $P_{6q}$  is the product of the  probability amplitudes for the 6-quark states to exist in the deuteron, and the factor of 2 enters because either state can be in the $d$-wave.
 Evaluation of $F_H$ using \eq{wf} leads to
 the result:
 \bea& b_1^{6q}(x)=-\sqrt{N_0N_2\over 2}{3\over4\pi}\int d^3p f_0f_2 (3\cos^2\theta-1)\delta\left({p\cos\theta+E(p)\over M}-x\right)P_{6q}.\nonumber\\&\label{b16}\eea
 
 To  proceed further we   specify the wave functions to be  harmonic oscillator wave functions. We  take $f_2(p)=-p^2R^2e^{-p^2R^2/2},\, f_0(p)=e^{-p^2R^2/2},$ where $R$ is the radius parameter.
Within the present framework, the model is specified by only three parameters-$R$, the quark mass $m$ and $P_{6q}$.  The key question is whether such a model can reproduce the HERMES data point at $x=0.452$ without using
a value of $P_{6q}$ large enough to conflict with conventional nuclear physics calculations that do not require a non-zero value. In other words, we ask if the hidden color states provide a substantial mechanism to make $b_1$ non-zero at large values of $x$. We note that the HERMES data 
has a large error bar, and our  numerical results  are chosen to reproduce the central value.

 Our procedure is to adjust the value of $P_{6q}$ to reproduce the data at that point and see how large 
a value is needed. To proceed further, we  use a quark mass of 338 MeV. We expect that the 6-quark state should be somewhat larger than that of a nucleon, and therefore choose
$R$ to be  1.2  fm. We shall see that the calculations are not very sensitive to the exact value of $R$. The  dependence on all three parameters  is examined below.

The evaluation of $b_1^{6q}(x)$ proceeds by using $d^3p=2\pi p^2dpd\cos\theta $, integrating over $\cos\theta$,  and changing variables to $u\equiv\,p^2R^2$.
The result is
\bea b_1^{6q}(x)={6MR \over\sqrt{30\pi}}\int_{u_{\rm min}(x)}^\infty\,du\,e^{-u}\left[3((x^2M^2+m^2)R^2+u-2xM R\sqrt{u+m^2R^2})-u\right]P_{6q},\label{b6q}\eea
where
\bea u_{{\rm min}(x)}\equiv {(x^2M^2-m^2)^2R^2\over 4x^2M^2}.\eea
\section{Results}

  \begin{figure}
  \centering
\includegraphics[width=8.991cm,height=10cm]{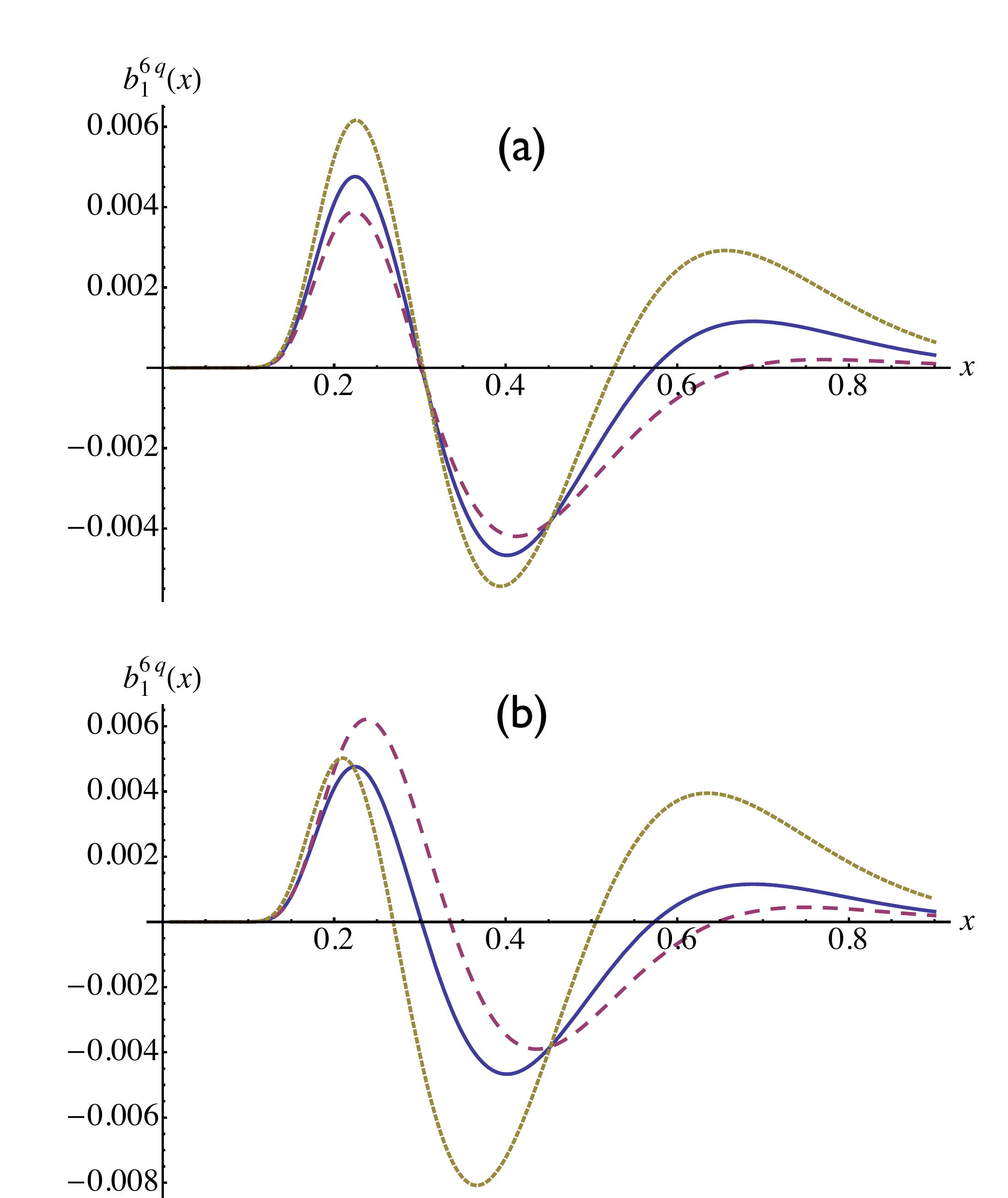}
\caption{(Color online) Computed values of $b_1^{6q}$ from \eq{b6q}. Sensitivity to parameters is displayed.  (a) Solid (blue) uses $R=1.2$ fm, m=338 MeV, long dashed (Red) $R$ is decreased by 10\%, dotted(green) $R$ is increased  by 10\%. (b) Solid (blue) uses $R=1.2$ fm, m=338 MeV, long dashed (Red) $m$ is increased by 10\%, dotted(green), $m$ is decreased  by 
10\%. This figure  is from Ref.~~\cite{Miller:2013hla}.} \label{fig:6q}\end{figure}

We now turn to the determine the contributions due to hidden color, $b_1^{6q}$ provided  by \eq{b6q}.
The value of $b_1^{6q}(x=0.452)$  is relevant because pionic and other  contributions are negligible, and  the measured value, $b_1=-3.8 \pm 0.16\times 10^{-3}$, differs from zero.
We choose  $P_{6q}=0.0015$ to reproduce the central value using $R=1.2$ fm and $m=338 $ MeV.  
The small value of $P_{6q}$  shows that the 6 quark configuration has great impact on the computed value  of $b_1$. It is noteworthy that  
such a very, very small value can not be ruled out by any observations. Our value of $P_{6q}$ is very small: small  enough
to evade any limits imposed by high accuracy conventional nuclear theory calculations of nucleon-nucleon scattering  and 
deuteron properties that  make no  explicit reference to hidden color states.  Our value of 0.0015 is  too small to conflict with conventional nuclear calculations.

The results for $b_1^{6q}$
 are shown in Fig.~\ref{fig:6q}.  Results using the  model   parameters $R=1.2$ fm, $m=338 $ MeV are shown as the solid curves in  Figs.~\ref{fig:6q}a and b, which also displays the sensitivity to the values of the parameters.
 There is relatively little sensitivity to the value of $R$, but more sensitivity to the value of $m$. However,
 there is  wider dependence upon the choice of models. For example, if the HERMES point at $x=0.452$ is not reproduced, this particular model will be ruled out.
 The exponential appearing in \eq{b6q} renders the contribution very small for small values of $x$. This is seen in the figure. However, there is a large negative contribution at values of $x\approx0.4$, as well as a double-node structure. The latter arises from the factor $3\cos^2\theta-1$ appearing in the integrand of \eq{b16}. The contributions of the hidden-color configurations are generally much smaller than those of exchanged pions except for values of $x$ larger than about 0.35. We also predict that, for even larger values of $x$,  $b_1$ changes sign and may have  another maximum. This mechanism allows contributions at large values of $x$. A quark in a hidden color, six quark configuration can have up to two units of $x$.
 The parameter dependence of the model is also explored. Fig.~\ref{fig:6q}a shows the dependence on the value of $R$ and 
Fig.~\ref{fig:6q}a shows the dependence on the value of $m$.  For each of the curves $P_{6q}$ is chosen so that the value at $x=0.452$ is the same. Shifting the value of $R$ while keeping $b_1^{6q}(0.452)$ fixed  requires less than 4\% changes in the value of $P_{6q}$. Increasing the value of the quark mass produces larger effects. Keeping $b_1^{6q}(0.452)$ fixed   requires that the value of
 $P_{6q}$ needs to be decreased by 20\% if the value of the quark  mass is increased by 10\%, and the value of  $P_{6q}$ needs to be increased by about a factor of 1.8 if the value of the quark  mass is decreased by 10\%. In the remainder of this paper, we use the central values $R=1.2$ fm, $m=338 $ MeV.

\begin{figure}
  \includegraphics[width=10.991cm,height=8cm]{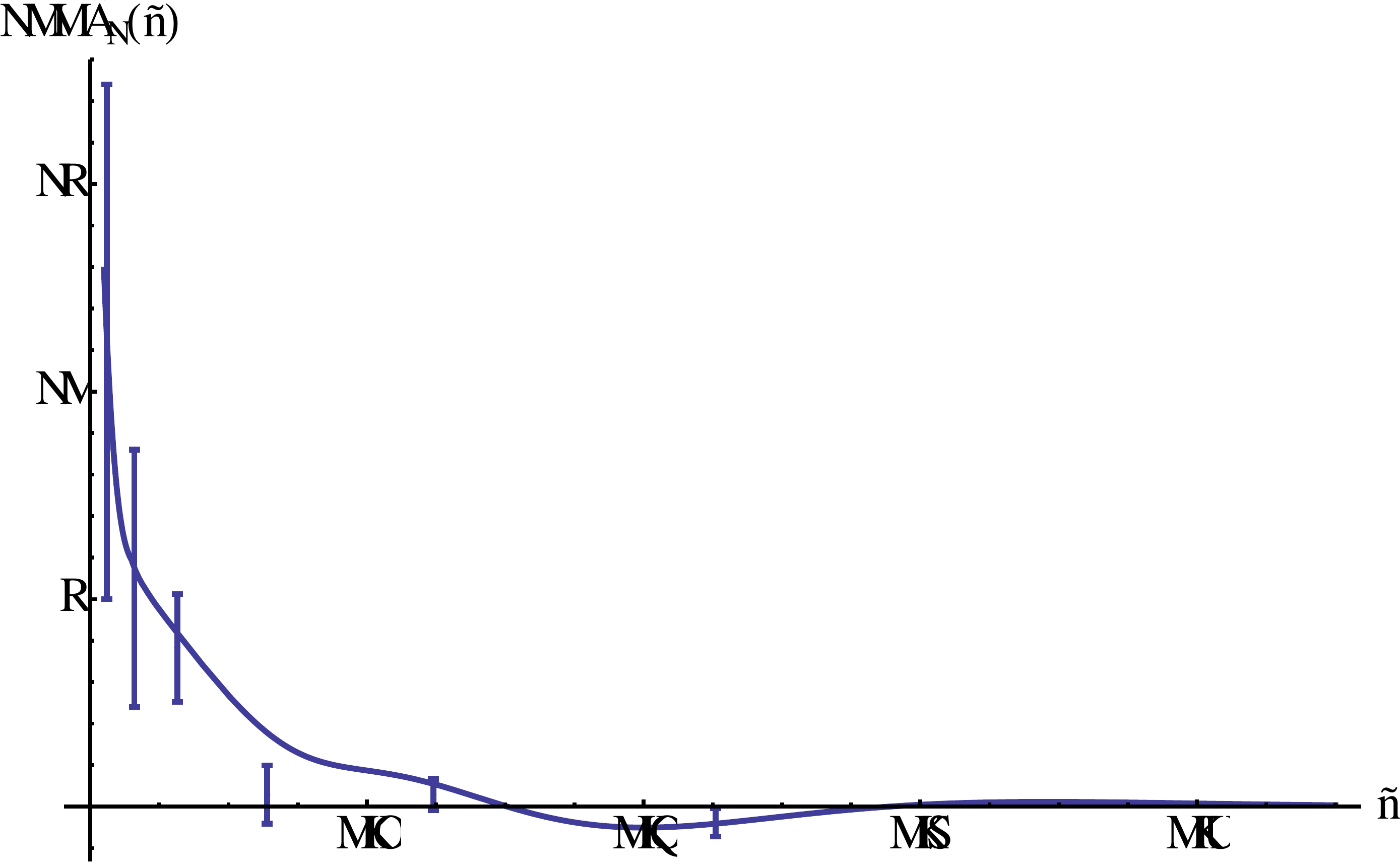}
\caption{ Computed values of $b_1=b_1^\pi+b_1^{6q}$.  The pion structure function is that of ~\cite{Aicher:2010cb},  model 1. This figure  is from Ref.~~\cite{Miller:2013hla}.}\label{HD}\end{figure}

 At this stage we can assess the size of our  computed $b_1^\pi$ and $b_1^{6q}$ versus the only existing data~\cite{Airapetian:2005cb}. These data is given in Table~\ref{b1result} along with our computed values of $b_1^\pi$ using the pion structure functions of Re.~\cite{Aicher:2010cb} and the three modes of~\cite{Sutton:1991ay}. These modes differ in the fraction of momentum carried by the sea: 10\%,15\% and 20\% for modes 1,and 3 respectively. The differences obtained by using different structure functions are generally not larger than the experimental error bars.
For values of  $x$ less than about  $0.2$, there is qualitative agreement between the measurements and the calculations of $b_1^\pi$ (which are much larger than those of $b_1^{6q}$), given  the stated  experimental uncertainties  and  the unquantifiable uncertainty caused by lack of knowledge of the sea. However, the large-magnitude negative central value measured at $x=0.452$ is two standard deviations away from the value provided by $b_1^{\pi}$ but in accord with the value provided by $b_1^{6q}$. 
Thus our result is that one can reproduce the Hermes measurements by using pion exchange contributions at low values of $x$ and hidden-color configurations at  larger values of $x$.  Thus our result is that one can reproduce the Hermes measurements by using pion exchange contributions at low values of $x$ and hidden-color configurations at  larger values of $x$. This is also shown in Fig.~\ref{HD}, where very good agreement between data and our model can be observed.
Please 
see~\cite{Miller:2013hla} for more figures and details.

\def\vum#1{\vspace*{#1cm}}
 \def\jum#1{\hspace*{#1cm}}
\begin{table}[t]  
\caption{ \label{b1result}
Measured values (in $10^{-2}$ units) of the  tensor structure function $b_1$. 
Both the  statistical and systematic uncertainties are listed.  The numbers in parenthesis refer to the structure function modes of Ref.~\cite{Sutton:1991ay} .}
\vum{-0.0}
\begin{tabular}{ccrrrrrrr} \hline\hline
$\langle x \rangle$ & $\langle Q^2\rangle$ &  
 $b_1$  &   $\pm\delta {b_1}^{\!\rm stat}$ &   $\pm\delta {b_1}^{\!\rm sys}$ & $\jum1 b_1^\pi$\cite{Aicher:2010cb}&$b_1^\pi$\cite{Sutton:1991ay} (1)&$b_1^\pi$\cite{Sutton:1991ay} (3) & $\quad b_1^{6q}$ \\
   & $\rm [GeV^2]$ &    $[10^{-2}]$ & $[10^{-2}]$ & $[10^{-2}]$ &\,\quad\hspace{.5cm}  $[10^{-2}]$ & $[10^{-2}]$  & $[10^{-2}]$&$\quad[10^{-2}]$\\
\hline
  0.012 &  0.51 &   11.20 &   5.51 &   2.77 &10.5 &15.5&24.1 &\quad0.00 \\
  0.032 &  1.06 &     5.50 &   2.53 &   1.84 &5.6 &6.8&8.9&0.00 \\
  0.063 &  1.65 &    3.82 &   1.11 &   0.60  &  4.2&3.7&4.1&0.00\\
  0.128 &  2.33 &      0.29 &   0.53 &   0.44 & 1.6& 1.3&1.3&0.01\\
  0.248 &  3.11 &    0.29 &   0.28 &   0.24 &-0.55& .13&0.12&0.41\\
  0.452 &  4.69 &    -0.38 &   0.16 &   0.03 &-0.02&-0.02&-0.022&-0.38\\ \hline\hline
\end{tabular}
\vum{-0.0}00
\end{table}

 \section{Summary}
 This paper contains an evaluation of the pion exchange and six-quark, hidden-color contribution to the $b_1$ structure function of the deuteron. The pion-nucleon form factor is constrained phenomenologically to 
 reduce a possible uncertainty. There is some numerical sensitivity to using different pionic structure functions. The pionic mechanism is sizable for small values of $x$, and can reproduce Hermes data~\cite{Airapetian:2005cb}  for values of $x$ less than 0.2. A postulated model involving hidden-color components of the deuteron is shown to
 complement the effects of pion exchange in reproducing the Hermes data for all measured values of $x$.  This model is based on the accuracy of the Hermes data for its largest value of $x=0.452$, and is chosen for simplicity. Many other models possible and we welcome further work to improve such models.   Nevertheless, the availability of the Hermes data enables us to make
  predictions for an upcoming JLab experiment~\cite{Jlab}. 
\begin{acknowledgements}
This material is based upon work supported by the U.S. Department of Energy Office of Science, Office of Basic Energy Sciences program under Award Number DE-FG02-97ER-41014.
\end{acknowledgements}

\bibliographystyle{spbasic}

\bibliography{b1bib}   

\begin{thebibliography}{21}
\providecommand{\natexlab}[1]{#1}
\providecommand{\url}[1]{{#1}}
\providecommand{\urlprefix}{URL }
\expandafter\ifx\csname urlstyle\endcsname\relax
  \providecommand{\doi}[1]{DOI~\discretionary{}{}{}#1}\else
  \providecommand{\doi}{DOI~\discretionary{}{}{}\begingroup
  \urlstyle{rm}\Url}\fi
\providecommand{\eprint}[2][]{\url{#2}}

\bibitem[{Aicher et~al(2010)Aicher, Schafer, and Vogelsang}]{Aicher:2010cb}
Aicher M, Schafer A, Vogelsang W (2010) {Soft-gluon resummation and the valence
  parton distribution function of the pion}. PhysRevLett 105:252,003,
  \eprint{1009.2481}

\bibitem[{Airapetian et~al(2005)}]{Airapetian:2005cb}
Airapetian A, et~al (2005) {First measurement of the tensor structure function
  b(1) of the deuteron}. PhysRevLett 95:242,001, \eprint{hep-ex/0506018}

\bibitem[{Arneodo(1994)}]{Arneodo:1992wf}
Arneodo M (1994) {Nuclear effects in structure functions}. PhysRept
  240:301--393

\bibitem[{Aubert et~al(1983)}]{Aubert:1983xm}
Aubert J, et~al (1983) {The ratio of the nucleon structure functions $F2_n$ for
  iron and deuterium}. PhysLett B123:275

\bibitem[{Bakker and Ji(2014)}]{Bakker:2014cua}
Bakker BL, Ji CR (2014) {Nuclear chromodynamics: Novel nuclear phenomena
  predicted by QCD}. ProgPartNuclPhys 74:1--34

\bibitem[{Bashkanov et~al(2013)Bashkanov, Brodsky, and
  Clement}]{Bashkanov:2013cla}
Bashkanov M, Brodsky SJ, Clement H (2013) {Novel Six-Quark Hidden-Color
  Dibaryon States in QCD}. PhysLett B727:438--442, \eprint{1308.6404}

\bibitem[{Carlson and Havens(1983)}]{Carlson:1983fs}
Carlson C, Havens T (1983) {Quark Distributions in Nuclei}. PhysRevLett 51:261

\bibitem[{Frankfurt and Strikman(1983)}]{frankfurt}
Frankfurt L, Strikman M (1983) {High Momentum Transfer Processes With Polarized
  Deuteronsv}. NuclPhys A405:557--580

\bibitem[{Geesaman et~al(1995)Geesaman, Saito, and Thomas}]{Geesaman:1995yd}
Geesaman DF, Saito K, Thomas AW (1995) {The nuclear EMC effect}.
  AnnRevNuclPartSci 45:337--390

\bibitem[{Guichon and Miller(1984)}]{Guichon:1983hi}
Guichon PA, Miller GA (1984) {Quarks and the Deuteron Asymptotic $D$ State}.
  PhysLett B134:15

\bibitem[{Harvey(1981)}]{Harvey:1988nk}
Harvey M (1981) {On the Fractional Parentage Expansions of Color Singlet Six
  Quark States in a Cluster Model}. NuclPhys A352:301

\bibitem[{Hen et~al(2013)Hen, Higinbotham, Miller, Piasetzky, and
  Weinstein}]{Hen:2013oha}
Hen O, Higinbotham D, Miller GA, Piasetzky E, Weinstein LB (2013) {The EMC
  Effect and High Momentum Nucleons in Nuclei}. IntJModPhys E22:1330,017,
  \eprint{1304.2813}

\bibitem[{Hoodbhoy et~al(1989)Hoodbhoy, Jaffe, and Manohar}]{jaffe}
Hoodbhoy P, Jaffe R, Manohar A (1989) {Novel Effects in Deep Inelastic
  Scattering from Spin 1 Hadrons}. NuclPhys B312:571

\bibitem[{Khan and Hoodbhoy(1991)}]{Khan:1991qk}
Khan H, Hoodbhoy P (1991) {Convenient parametrization for deep inelastic
  structure functions of the deuteron}. PhysRev C44:1219--1222

\bibitem[{Koch and Miller(1985)}]{Koch:1985y}
Koch V, Miller G (1985) {SIX QUARK CLUSTER EFFECTS AND BINDING ENERGY
  DIFFERENCES BETWEEN MIRROR NUCLEI}. PhysRev C31:602--612

\bibitem[{Miller(1984)}]{Miller:1985un}
Miller GA (1984) {SIX QUARK CLUSTER COMPONENTS OF NUCLEAR WAVE FUNCTIONS AND
  THE PION NUCLEUS DOUBLE CHARGE EXCHANGE REACTION}. PhysRevLett 53:2008--2011

\bibitem[{Miller(2014)}]{Miller:2013hla}
Miller GA (2014) {Pionic and Hidden-Color, Six-Quark Contributions to the
  Deuteron b1 Structure Function}. PhysRev C89:045,203, \eprint{1311.4561}

\bibitem[{Miller and Gal(1987)}]{Miller:1987ku}
Miller GA, Gal A (1987) {Quark Model of the $\pi^- P P \to P N$ Reaction}.
  PhysRev C36:2450--2458

\bibitem[{Slifer and Long(2013)}]{Jlab}
Slifer K, Long E (2013) {Novel Physics with tensor polarized deuteron targets}.
  PoS PSTP2013:008, \eprint{1311.4835}

\bibitem[{Sutton et~al(1992)Sutton, Martin, Roberts, and
  Stirling}]{Sutton:1991ay}
Sutton P, Martin AD, Roberts R, Stirling WJ (1992) {Parton distributions for
  the pion extracted from Drell-Yan and prompt photon experiments}. PhysRev
  D45:2349--2359

\bibitem[{Umnikov(1997)}]{Umnikov:1996qv}
Umnikov AY (1997) {Relativistic calculation of structure functions b(1,2)(x) of
  the deuteron}. PhysLett B391:177--184, \eprint{hep-ph/9605291}

\end{thebibliography}


\end{document}